\documentclass[prl,twocolumn,showpacs,superscriptaddress,preprintnumbers,amsmath,amssymb]{revtex4-1}

\usepackage{graphicx}
\usepackage{dcolumn}
\usepackage{bm}
\usepackage{hyperref}
\usepackage{amsmath}
\usepackage{amssymb}
\usepackage{subfigure}
\usepackage{epstopdf}
\usepackage{color}
\usepackage{wasysym}

\begin{document}

\begin{abstract}
  We study persistent currents for interacting one-dimensional bosons on a tight ring trap, subjected
  to a rotating barrier potential, which induces an artificial $U(1)$ gauge field.
  We show that, at intermediate interactions, the persistent current response is maximal,
  due to a subtle interplay of effects due to the barrier, the interaction and quantum
  fluctuations.
  These results are relevant for ongoing experiments with ultracold atomic gases
  on mesoscopic rings.
\end{abstract}



\title{Optimal Persistent Currents for Interacting Bosons on a Ring with a Gauge Field}

\author{Marco Cominotti}
\affiliation{Universit\'e Grenoble Alpes, LPMMC, F-38000 Grenoble, France}
\affiliation{CNRS, LPMMC, F-38000 Grenoble, France}

\author{Davide Rossini}
\affiliation{NEST, Scuola Normale Superiore and Istituto Nanoscienze-CNR, I-56126 Pisa, Italy}

\author{Matteo Rizzi}
\affiliation{Institut f\"ur Physik, Johannes Gutenberg-Universit\"at Mainz,Staudingerweg 7, D-55099 Mainz, Germany}

\author{Frank Hekking}
\affiliation{Universit\'e Grenoble Alpes, LPMMC, F-38000 Grenoble, France}
\affiliation{CNRS, LPMMC, F-38000 Grenoble, France}

\author{Anna Minguzzi}
\affiliation{Universit\'e Grenoble Alpes, LPMMC, F-38000 Grenoble, France}
\affiliation{CNRS, LPMMC, F-38000 Grenoble, France}

\maketitle

A quantum fluid confined on a ring and subjected to a $U(1)$ gauge potential
displays a periodicity in the particle current as a function of the flux
of the corresponding classical gauge field. This persistent current phenomenon
is a manifestation of the Aharonov-Bohm effect, and reflects the macroscopic
coherence of the many-body wave function along the ring.
Such currents were observed more than 50 years ago in bulk superconductors~\cite{Deaver}
and, more recently, in normal metallic rings, overcoming the challenges
of the decoherence induced by inelastic scattering~\cite{Levy}.
The most recent developments in the manipulation of ultracold atoms on ring
traps~\cite{gupta, nist} have disclosed a novel platform for the study of persistent currents,
which can be induced by the application of a rotating localized barrier
or, alternatively, by inducing suitable artificial gauge fields~\cite{dalibard}.
Tunable localized barriers in toroidal Bose-Einstein condensates have been realized, using well-focused, repulsively tuned laser beams~\cite{nist}. Also, recently the engineering of an atomic superconducting quantum interference device was demonstrated~\cite{ryu}.
The unprecedented variety of interaction and barrier strength regimes paves the way
to applications such as high-precision measurements, atom interferometry and quantum information,
{\it e.g.}, by the construction of macroscopic superposition of current states
and flux qubits~\cite{hallwood, schenke, amico}.

The scenario becomes particularly intriguing if the transverse section of the ring
is sufficiently thin to effectively confine the system in one dimension (1D):
the rich interplay between interactions, quantum fluctuations, and statistics
acquires a role of primary relevance.
In absence of any obstacle along the ring, the persistent currents display
an ideal sawtooth behavior as a function of the flux, {\it i.e.},
perfect superfluidity for any interaction strength at zero temperature~\cite{leggett,loss}.
Diamagnetic or paramagnetic response depending on the population parity
is expected~\cite{Zvyagin} for fermions but not for bosons.
If a localized barrier is added, persistent currents are smeared---their shape
taking a sinusoidal form in the case of large-barrier or small-tunneling limit---as obtained
for thin superconducting rings from a Luttinger-liquid approach~\cite{hekking}.
Beyond these limiting regimes the physics of bosonic persistent current remains unexplored.

The aim of the present work is to provide a complete characterization of persistent currents
for 1D bosons, in all interaction and barrier strength regimes.
By combining analytical as well as numerical techniques suited for the 1D problem,
we show that the current amplitude is a nonmonotonic function of the interaction strength
and displays a pronounced maximum in all regimes of barrier height.
The presence of an optimal regime illustrates the highly nontrivial
combination of correlations, quantum fluctuations and barrier effects.
Our results demonstrate that, in a large range of interaction strengths,
unwanted impurities or imperfections on the ring only weakly affect the system properties.
For the application to quantum state manipulation, the regimes of choice
should be either very weak or very strong interactions, where the response
to a localized external probe is stronger. Our predictions are readily amenable
to experimental testing with quasi-1D ultracold atomic gases confined
in mesoscopic uniform and lattice rings.
Indeed all the interaction regimes, from weakly interacting quasicondensate
to the impenetrable-boson Tonks-Girardeau limit, have been experimentally demonstrated
for 1D bosonic wires~\cite{dettmer}.

{\em Persistent currents for bosons under a gauge field.}---Let us consider
a 1D ring of circumference $L$, containing $N$ bosons
of mass $M$ interacting with each other via a contact
potential $v(x-x')=g \, \delta(x-x')$. The ring contains a localized barrier modeled as
$U_{b}(x,t)=U_{0}\delta(x-Vt)$, moving along its
circumference at constant velocity $V$.
This induces an effective gauge field with Coriolis flux
$\Omega = MVL/2\pi\hbar$ in the rotating frame.
In the corotating frame, the Hamiltonian reads
\begin{equation}
  \mathcal{H} \! = \!\! \sum_{j=1}^{N} \frac{\hbar^2}{2M}
  \! \left( \!\! -i\frac{\partial}{\partial x_{j}} - \frac{2\pi}{L}\Omega \! \right)^{\!\! 2} \!
  + U_0\delta(x_j) + \frac g 2 \! \sum_{j,l=1}^{N}\delta(x_{l}-x_{j}) \,.
  \label{eq:hamiltonian}
\end{equation}
This generalizes the Lieb-Liniger model~\cite{liebliniger} to the rotating case
and is nonintegrable due to the presence of the barrier.
The corresponding many-body energy spectrum is periodic in $\Omega$ with period 1, giving rise to the Aharanov-Bohm effect.

We consider the stationary regime, reached after the barrier
has been adiabatically switched on at early times,
such that no high-energy excitations are created.
At zero temperature, the spatially averaged particle current $I(\Omega)$, or persistent current, is obtained from the ground-state energy $E(\Omega)$ via the thermodynamic relation~\cite{Bloch1970}
\begin{equation}
  I(\Omega)=-\frac{1}{2\pi\hbar}\frac{\partial E(\Omega)}{\partial \Omega}\,.
  \label{eq:current}
\end{equation}

\begin{figure}[!t]
  \centering
  \includegraphics[width=0.254\textwidth]{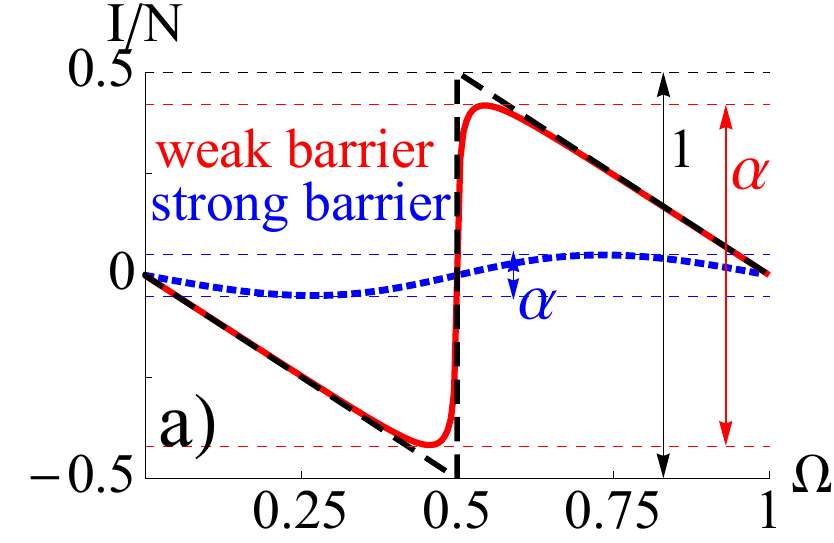}\includegraphics[width=0.245\textwidth]{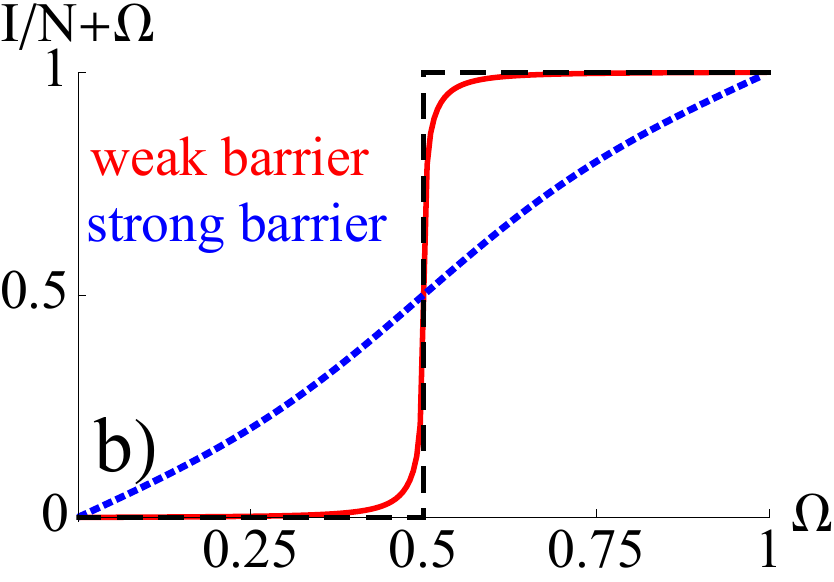}
  \caption{Persistent current
    (a) in the rotating frame and (b) in the nonrotating frame,
    in units of $I_{0}=2\pi\hbar/ML^2$ for $N=18$, for zero  barrier (black-dashed line), weak barrier ($\lambda=0.2$, red-solid line) and strong barrier ($\lambda=10$, blue dotted line) at fixed interaction strength ($\gamma=0.004$).}
  \label{fig:currents}
\end{figure}

In absence of a barrier, for any interaction strength, the ground-state energy  is given by a series of parabolas with well-defined angular momentum,
shifted with respect to each other by Galilean translation~\cite{Bloch1973}
and intersecting at the frustration points $\Omega_j = (2 j +1)/2$.
The corresponding persistent current (Fig.~\ref{fig:currents}) is a perfect
sawtooth in the rotating frame and a staircase in the nonrotating frame,
corresponding to states with well-defined angular momentum (dashed lines).
The addition of a  barrier breaks the rotational invariance and induces coherent superpositions of states of different angular momentum. This gives rise to gap openings in the many-body energy spectrum at the frustration points, thus forming the working points for a qubit based on a superposition of current states. The level mixing is visible in the persistent current. In the rotating frame, $I(\Omega)$ is a smeared sawtooth for weak barriers  (solid lines) and a sinusoid for large barriers (dotted lines).  In order to characterize the qubit, we focus  on the amplitude  $\alpha$ of the persistent current in the rotating frame, determining its value for all regimes of dimensionless interaction
strength $\gamma=Mg/\hbar^2n_{0}$, with $n_{0}=N/L$,
and dimensionless barrier height $\lambda=MU_{0}L/\pi\hbar^2$.

\begin{figure*}[t]
  \includegraphics[width=0.243\textwidth]{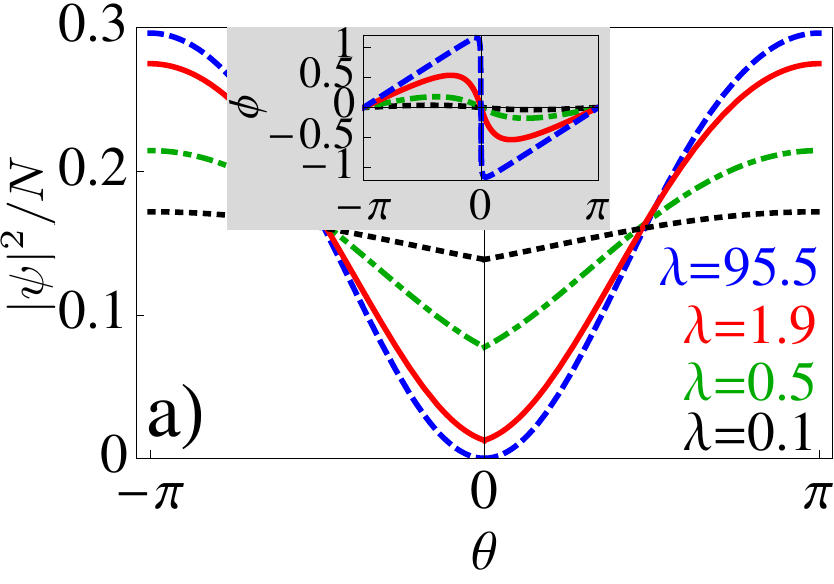}
  \hspace{1mm}
  \includegraphics[width=0.23\textwidth]{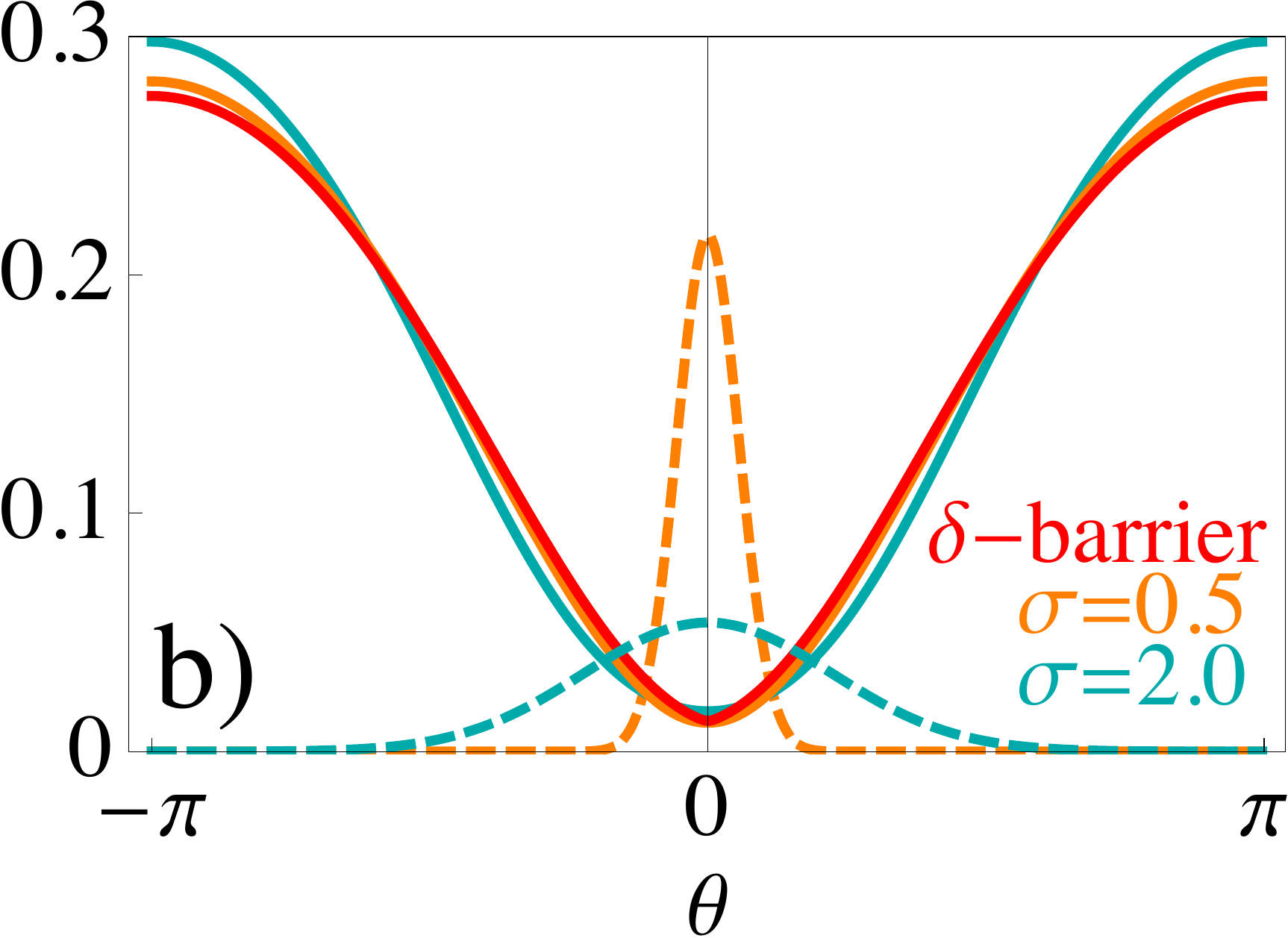}
  \hspace{1mm}
  \includegraphics[width=0.235\textwidth]{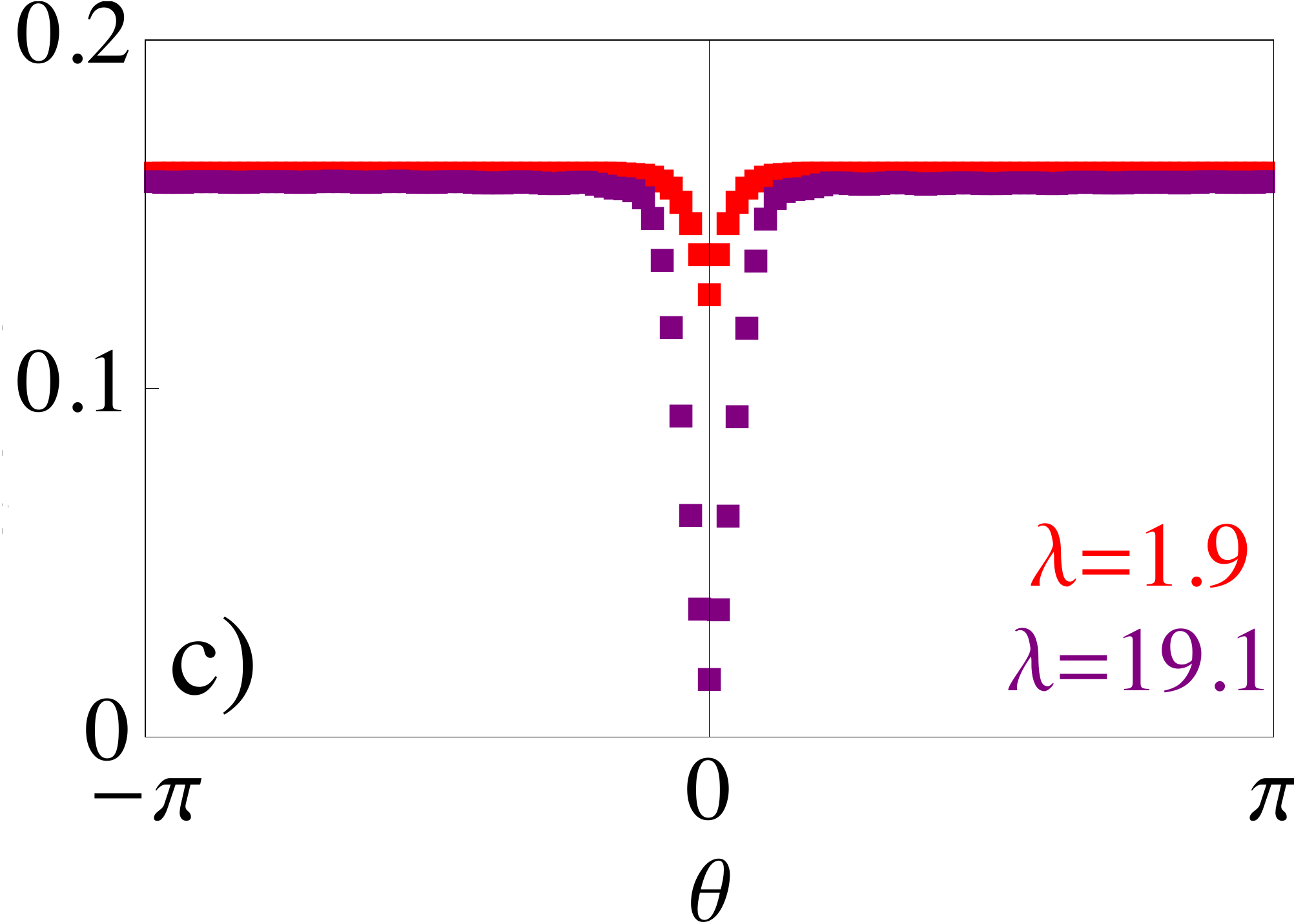}
   \hspace{1mm}
  \includegraphics[width=0.235\textwidth]{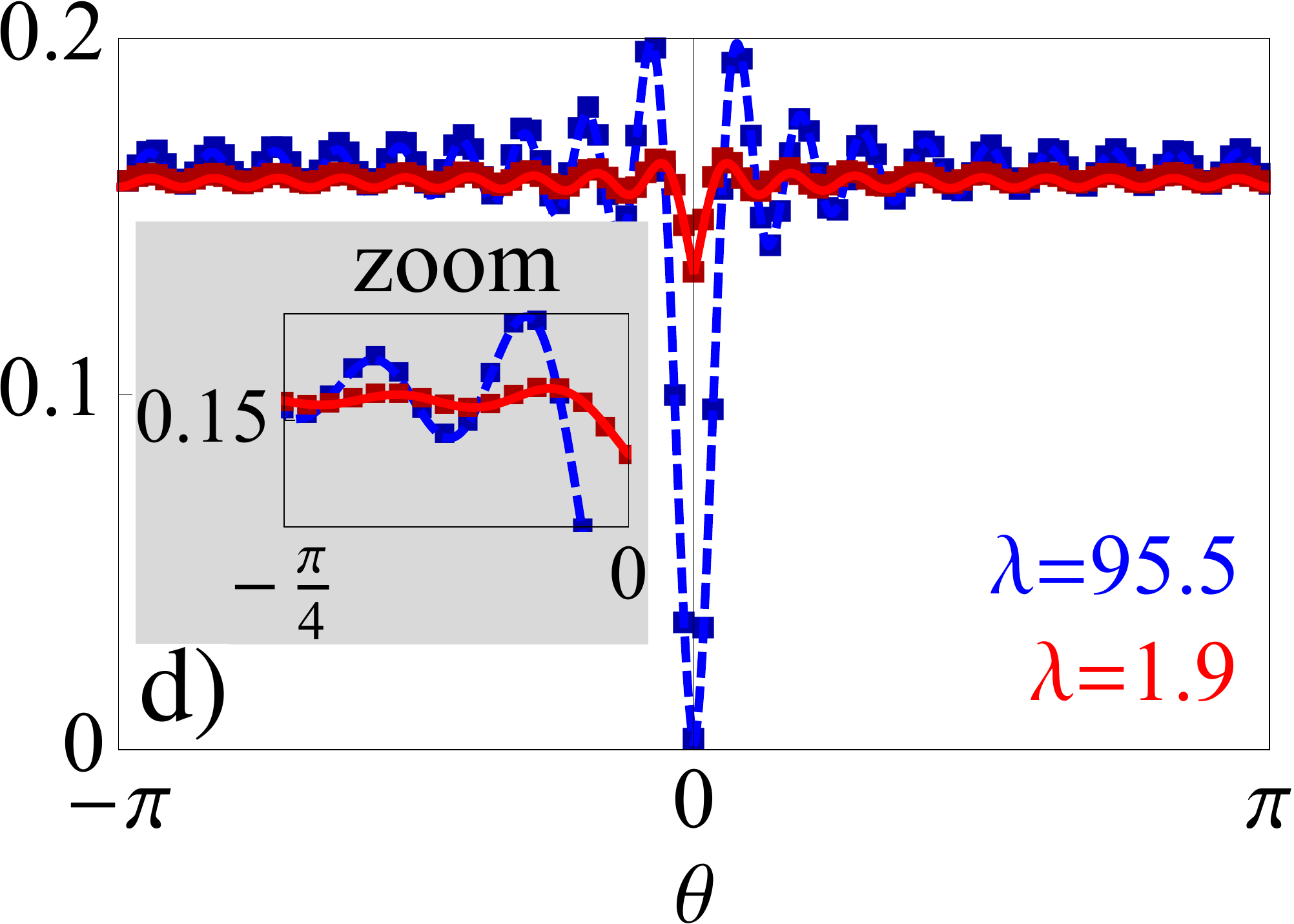}
  \caption{Density profiles vs angular coordinate $\theta$ along the ring.  (a) GP analytical solution at weak interaction strength $\gamma=0.01$ for a delta barrier for various values of the barrier strength $\lambda$; the inset shows the soliton phase vs $\theta$ for the same values of barrier strength. (b) GP numerical solution (solid lines) at $\gamma=0.01$ for a Gaussian barrier of strength  $\lambda=1.9$  for various values of the barrier width $\sigma$  (in units of $n_0^{-1}$), and corresponding  barrier potential (dashed lines, in units of $10\times2\pi^{2}\hbar^2/mL^{2}$). (c) MPS solution at intermediate interaction $\gamma=3.3$ and filling $0.15$. (d) Analytical TG solution (lines) vs MPS (squares) in the hard-core limit.
    The other parameters used are $N=18$ and $\Omega=0.4$ in all the curves.}
  \label{fig:density}
\end{figure*}

{\em Noninteracting and impenetrable boson limits.}---Both for zero and infinitely large repulsive interactions,  it is
possible to find an exact solution to the many-body Schr\"odinger equation
$\mathcal{H}\Psi(x_{1},...\, x_{N})=E \Psi(x_{1},...\, x_{N})$.
For a noninteracting (NI)
Bose gas, the many-body wave function $\Psi_{\rm NI}(x_{1},...\,
x_{N})=\prod_{i=1}^{N}\psi_0(x_{i})$ is simply given by the product of $N$ identical
single-particle wave functions $\psi_0(x_{i})$, which are ground-state solutions of the
corresponding one-body Schr\"odinger equation, $ \frac{\hbar^2}{2M}\left(-i \partial_x -
\frac{2\pi}{L}\Omega\right)^2 \psi_n +U_0 \delta(x) \psi_n = \varepsilon_n \psi_n$, and has energy
$E_{\rm NI}= N \varepsilon_0$ . In the infinitely interacting limit of impenetrable
bosons, or Tonks-Girardeau (TG) gas, the solution is obtained by mapping the system onto a gas of
non-interacting fermions subjected to the same external potential~\cite{girardeau},
$\Psi_{\rm TG}(x_{1},...\, x_{N})= \prod_{1 \leq j<\ell \leq N} {\rm sgn}(x_j-x_\ell)
\det[\psi_k(x_i)]$. The corresponding energy $E_{\rm TG}=\sum_{k=0}^{N-1} \varepsilon_k$
and density profile $n(x)=\sum_{k=0}^{N-1} |\psi_k|^2$ directly reflect the fermionization properties of the strongly interacting  Bose gas. The Friedel oscillations of the density profile [Fig.~\ref{fig:density}(d)] are indeed  a signature of the strongly correlated regime~\cite{friedel}.

The persistent current amplitude in the zero and infinitely interacting limits, obtained from Eq.~\eqref{eq:current} (see the Supplemental Material ~\cite{SM}), is shown in Fig.~\ref{fig:alpha}. We find that the current amplitude depends on the interaction regime. In particular, for all values of barrier strength,  $\alpha$ is  always larger in the strongly interacting regime than in the noninteracting one. This behavior is explained by noticing that the barrier affects the lowest-lying energy levels more than the high-energy ones~\cite{schenke}. As a result, the current amplitude  is smaller for noninteracting bosons, occupying only the lowest level, than for the TG gas, where the levels are filled up to the Fermi energy.
%

{\em Weak interactions.}---In order to explore the role of interactions for quantum state manipulation, we start from the noninteracting result and consider the effect of weakly repulsive interactions. In this regime we neglect quantum fluctuations~\cite{petrov} and describe the fluid as a Bose-Einstein condensate, making use of the mean-field
Gross-Pitaevskii (GP) equation. In the corotating frame, this takes the form
$ \frac{\hbar^2}{2M} ( -i\partial_x - \frac{2\pi}{L}\Omega )^2 \Phi + U_0\delta(x) \Phi + g|\Phi|^2 \Phi = \mu\Phi$,
where $\Phi$ is the condensate wave function and $\mu$ the chemical potential.
This equation admits stable dark-soliton  solutions, with a size given by the healing length $\xi=\hbar/\sqrt{2M g n_0}$. 
We have found an analytical solution for $\Phi$ in terms of Jacobi
elliptic functions~\cite{SM}, thus extending Refs.~\cite{kanamoto}
[see also Ref.~\cite{kamenev}]. In particular, by comparing it
with the numerical solution of the GP equation, we find that the barrier pins the soliton and turns it to the ground-state solution.  The resulting density profile has a minimum at the barrier position. The depth and the healing length  of the soliton depend on the interaction strength, the barrier height, and the rotation velocity [see Fig.~\ref{fig:density}(a) and~\cite{SM}]. The solitary suppression of the density is accompanied by a winding of the superfluid phase, known as a phase-slip [inset of Fig.~\ref{fig:density}(a)].  We note that a finite-width barrier, as could be realized experimentally,  yields very similar profiles [Fig~\ref{fig:density}(b)]. Its extension to multiple finite-width barriers  is known to give rise to a wealth of nonlinear modes \cite{malomed}. 

The soliton energy is obtained from the GP energy
functional $E_{\rm GP}[\Phi]=\int \mbox{d}x\,\Phi^* \hbar^{2}/2M(-i \partial_x - \frac{2\pi}{L}\Omega)^2\Phi + g
|\Phi|^4/2 + U_0\delta(x) |\Phi|^2$, which encodes the dependence on $\Omega$ also in the condensate wave function. Computing the persistent current amplitude $\alpha$ we find that it increases monotonically with the
interaction strength $\gamma$, as illustrated in Fig.~\ref{fig:alpha}.
This is due to the fact that the healing length $\xi$ decreases at increasing interaction strength, thus yielding a more effective screening of the barrier, thereby restoring superfluidity.

{\em Strong interactions.}--- We now turn to the effect of quantum
fluctuations on the persistent current at strong interparticle interactions, up to the TG limit.
This can be achieved with the help of Luttinger liquid (LL) theory~\cite{haldane},
a low-energy, quantum hydrodynamics description of the bosonic fluid,
in terms of the canonically conjugate fields $\theta$ and $\phi$ corresponding to
density fluctuations and superfluid phase, respectively.
In the rotating frame, the effective LL Hamiltonian for a uniform ring is
\begin{equation}
  \mathcal{H}_{0}=\frac{\hbar v_{s}}{2\pi}\!\!\int_{0}^{L}\!\!\!\mbox{d}x\bigg[K\!\!\left(\partial_{x}\phi(x)\!-\!\frac{2\pi}{L}\Omega \right)^2\!\!\!+\!\frac{1}{K}(\partial_{x}\theta(x))^2 \bigg]\,.
  \label{eq:llhamiltonian}
\end{equation}
The microscopic interaction strength enters through
the Luttinger parameter $K$ and the sound velocity $v_s$.
In the case of repulsive contact interactions their dependence on the interaction
strength is known (see, {\it e.g.}, Ref.~\cite{cazalilla}):
at vanishing interactions, $K$ tends to infinity and $v_s$ vanishes,
while, in the TG limit, $K=1$ and $v_s$ corresponds to the Fermi velocity of the fermionized Bose gas. 

In the presence of a barrier, nonlinear terms appear in the Hamiltonian. We  treat these perturbatively in the limits of weak and strong barrier strength. In  the regime of a {\em weak barrier}, its contribution to the Hamiltonian $\mathcal{H}_{b}=\int \mbox{d}x\, U_0 \delta(x) \rho(x)$
is obtained by keeping only the lowest harmonics in the density field expansion
$\rho(x)=[n_{0}+\partial_{x}\theta(x)/\pi]\sum_{l=-\infty}^{+\infty}e^{2il\theta(x)+2il\pi n_{0}x}$,
yielding as the most important term $\mathcal{H}_{b}\sim 2 U_{0}n_{0}\cos[2\theta(0)]$.
This backscattering term breaks angular momentum conservation,
allowing for transitions between angular momentum states~\cite{citro}. At the same time, quantum fluctuations renormalize the barrier strength down according 
to $U_{\rm eff}=U_0 (d/L)^K$,
where $d$ is a suitably chosen short-distance cutoff~\cite{SM}.
As a result, the persistent current
close to the frustration point $\Omega=1/2$ is a smeared sawtooth
\begin{equation}
  I(\Omega)/N=-I_{0} \, \delta\Omega \Big[ 1-\Big( \sqrt{4\delta\Omega^{2}+\lambda_{\rm eff}^{2}/\pi^{2}} \Big)^{-1} \Big]\,,
\end{equation}
where $\delta\Omega=\Omega-1/2$, $\lambda_{\rm eff} = MU_{\rm eff}L/\pi\hbar^{2}$,
and $I_0=2\pi\hbar/ML^2$. The corresponding amplitude $\alpha$ is shown
in Fig.~\ref{fig:alpha}. For decreasing interactions down from the TG limit, the quantum fluctuations of density increase, 
suppressing
the barrier more strongly, hence,  
increasing $\alpha$.

In the opposite case of {\em strong barrier strength}, we model the transport
across the barrier with a tunneling Hamiltonian
$\mathcal{H}_{t}=- 2 t n_0\cos[\phi(L)-\phi(0)+2\pi\Omega]$, where $t$ is
the tunneling amplitude. In this limit, quantum phase fluctuations lead to a renormalization down of the tunnel
amplitude~\cite{renormalization}, i.e.,
\begin{equation}
  I(\Omega)/N=-(2t/L\hbar)(d/L)^{1/K}\sin(2\pi\Omega)\,.
\end{equation}
Note the dual nature of this result (power $1/K$) in comparison to the one obtained above for a weak barrier (power $K$). Indeed, the quantum phase fluctuations increase with increasing interactions, thereby suppressing the tunneling amplitude more, thus, decreasing $\alpha$. The duality between the two models can also be used to establish a link between the tunnel amplitude and the barrier height~\cite{weiss,SM}.
The persistent current amplitude for large barrier is also shown in Fig.~\ref{fig:alpha}.
Interestingly, the renormalization by quantum fluctuations at intermediate interactions
is effective enough to turn a relatively large barrier into a weak one. This quantum healing phenomenon completely changes the physical scenario and illustrates the dramatic effect of interplay of interactions and quantum fluctuations. 

Both for weak and large barriers, the LL description breaks down for sufficiently weak interactions when the short-distance cutoff required in the theory increases, until it becomes comparable with the system
size.

\begin{figure}[t]
  \centering
\includegraphics[width=0.42\textwidth]{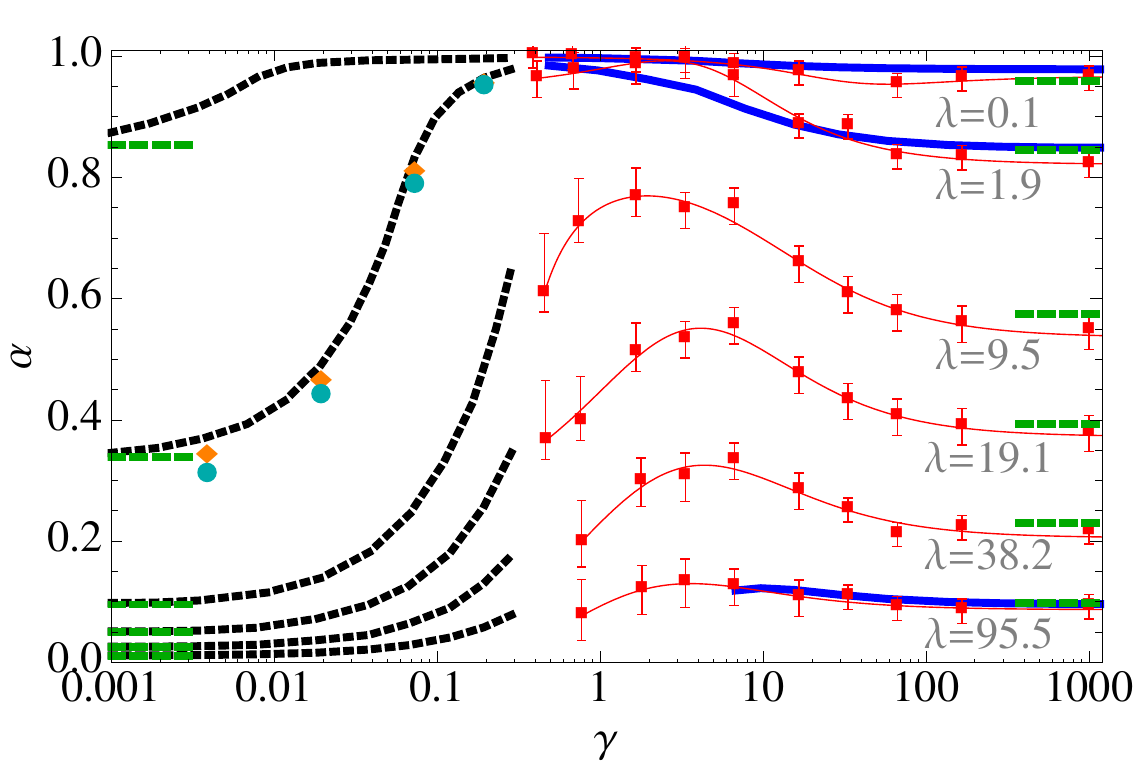}
  \caption{Persistent current amplitude $\alpha$, in units of $I_{0}=2\pi\hbar/ML^2$,
    as a function of the interaction strength $\gamma$
    at varying  barrier strength $\lambda$, for $N=18$, 
    from GP equation (black dotted lines),
    LL approach (blue solid lines), 
    numerical MPS calculations (red squares, red thin lines are guides to the eye) and NI and TG exact solutions (green dashes). Orange diamonds and cyan circles are for a Gaussian barrier of width $\sigma=0.5, 2 \,(n_{0}^{-1})$  respectively.}
  \label{fig:alpha}
\end{figure}

{\em Intermediate interactions and barrier strengths.}---Away from the weakly and the
strongly interacting regime and for arbitrary barrier strength it is difficult to tackle
the many-body Schr\"odinger equation corresponding to the Hamiltonian of Eq.~\eqref{eq:hamiltonian}
with analytical approaches; therefore, we employ numerical simulations based on the
density-matrix renormalization-group (DMRG)~\cite{schollwock}.
After discretizing the space on $N_s$ sites, and using the standard Peierls construction,
we map Eq.~\eqref{eq:hamiltonian} onto a Bose-Hubbard model on a 1D ring lattice at low filling:
%
\begin{eqnarray}
  \nonumber
  \mathcal{H}_{\rm lat} & \! = & \!
  \sum_{j=1}^{N_s} \! -t_{\rm BH}\left( \! e^{-\frac{2 \pi i \Omega}{N_s}} b^\dagger_j b_{j+1} \! + \! {\rm H.c.} \! \right) \!
  + \! \frac{U_{\rm BH}}{2} \! n_j (n_j \! - \! 1) \\
  & & + \left( \beta \, \delta_{j,1} n_{j} - \mu \, n_j \right) \,,
  \label{eq:BH_model}
\end{eqnarray}
where $b^\dagger_j$ ($b_j$) are bosonic creation (annihilation) operators at site $j$,
$t_{\rm BH}$ is the tunnel energy for bosons on adjacent sites,
and $U_{\rm BH}$ is the on-site repulsion energy.
The phase twist $\Omega / N_s$ accounts for the effect of the Coriolis flux,
$\beta$ is the barrier strength, and $\mu$ is the chemical potential.
An experimental implementation of the Hamiltonian of Eq.~\eqref{eq:BH_model} with a localized barrier
on one lattice site has been recently proposed~\cite{amico}.
We compute the ground-state wave function and energy
using a matrix product state (MPS)~\cite{danshita}, optimized for periodic boundary conditions~\cite{SM}.
The resulting density profiles are shown in Fig.~\ref{fig:density} (c, d).
As compared to  Figs.~\ref{fig:density} (a), the healing effect is much more pronounced and might be efficiently used to screen unwanted impurities.
As anticipated above, for large interactions the density profiles display Friedel oscillations,  in very good agreement with the analytical predictions in the TG limit. The oscillations are strongly damped at intermediate and weak interactions. We also compute the amplitude $\alpha$
of the persistent current for a large range of barrier heights $\lambda$
and interaction strengths $\gamma$ (Fig.~\ref{fig:alpha})~\cite{mps}.
We find a nonmonotonic behavior connecting the weak- and strong-interaction
results as a consequence of the subtle interplay between backscattering and interaction effects. 
This result allows us to confirm the expected regimes of validity of the analytical estimates.
We  note that the position of the maximum, i.e.,
the optimum persistent current, depends on the barrier strength.

{\em Experimental considerations.}--- Our results are also valid for finite-width barriers, used in realistic implementations. We have checked that the use of a Gaussian barrier  does not lead to qualitative changes of our results in the mean-field regime~\cite{discrepancy} (see  Fig.  \ref{fig:alpha} for a choice of barrier strength; similar results hold for all the barrier strength regimes), provided that the barrier width remains comparable to the interparticle distance, a condition achievable e.g., with a microscope-focused stirring beam~\cite{Desbuquois}. Thermal fluctuations give rise to additional smearing of persistent currents starting from a typical temperature $k_BT\sim N E_{0}= 2\pi^2\hbar^2 n_0/ML$~\cite{loss,Zvyagin,das}. Within typical experimental constraints on energy and time scales, persistent currents might be observed in the next-generation experiments of mesoscopic ring traps on a chip
({\it i.e.}, with a ring diameter of $\sim 5 \;\mu$m),
corresponding to a typical energy $N E_0\sim 550$ Hz for $^{87}$Rb atoms taking $N=18$, as in Fig.~\ref{fig:alpha}.

{\em Conclusions.}---We disclosed the role of quantum fluctuations and correlations
in the emergence of persistent currents of rotating bosons in reduced dimensionality,
in the presence of a localized barrier.
Our results, summarized in Fig.~\ref{fig:alpha}, evidence the presence
of a maximum as a function of the interaction strength,
which is due to the competition between correlations and fluctuations.
While at increasing interactions a classical bosonic field screens the barrier more
and more, going towards the strongly correlated Tonks-Girardeau regime
quantum fluctuations screen the barrier less and less, due to the increasingly fast spatial decay of phase-phase correlations~\cite{bcs}.
These predictions can be tested via time-of-flight measurements,
similar to those used to probe circulation in atomic rings~\cite{gupta,nist}.

In the present work we studied a thermodynamic quantity---the persistent current---but the interplay of barrier and quantum fluctuations will have a similar impact on out-of-equilibrium properties involving wave emission, ranging from collective modes to transport phenomena, including shock wave generation
and Hawking radiation of superfluids~\cite{pavloff}.
Our results are also relevant to other mesoscopic bosonic quantum fluids,
as thin superconducting rings, photons in nonlinear twisted-pipe waveguides,
and solid-state photonic or polaritonic nanocavities etched on a ring-necklace shape.

{\it Acknowledgments.} We are indebted to L. Amico, R. Citro, R. Fazio,
L. Glazman, N. Pavloff, I. Safi, P. Silvi and W. Zwerger for discussions.
This work is supported by the ERC Handy-Q grant
N.258608,  Institut universitaire de France,  Italian MIUR through FIRB
Project RBFR12NLNA. MPS simulations were performed on the TQO cluster of the
Max-Planck-Institut f\"ur Quantenoptik (Garching) and on MOGON cluster in Mainz.

\newpage

\appendix

\onecolumngrid

\begin{center}
{\large Supplemental Material for: \\
``Optimal Persistent Currents for Interacting Bosons on a Ring with a Gauge Field''}
\end{center}

\section{Details of the exact solutions for the limiting cases of ideal and Tonks-Girardeau gas}

We outline the solution for the single particle problem used to obtain the
ideal-gas and the Tonks-Girardeau (TG) solution of the main text. The single particle
eigenfunctions of the one-body Schr\"odinger equation take the form 
\begin{equation}
  \psi_{n}(x; \Omega) = \left \{
  \begin{array}{ll}
    \frac{1}{\mathcal{N}_n} \, e^{-i\Omega \pi} \left[ e^{ik_{n}(x-L/2)}+A_{n,\Omega}e^{-ik_{n}(x-L/2)} \right] & x\in[0,L/2)
      \vspace*{2mm} \\
    \frac{1}{\mathcal{N}_n} \, e^{i\Omega \pi} \left[ e^{ik_{n}(x+L/2)}+A_{n,\Omega}e^{-ik_{n}(x+L/2)} \right] & x\in[-L/2,0) \\
  \end{array}
  \right.
\end{equation}
By imposing twisted boundary conditions, unity normalization and the cusp condition
$\partial_{x}\psi_{n}^{+}(0^+; \Omega)-\partial_{x}\psi_{n}^{-}(0^-; \Omega)=\lambda\psi_{n}(0; \Omega)$,
we obtain 
$$
k_{n}=\pm\lambda\frac{\pi}{L}\frac{\sin(k_{n}L)}{\cos(2\pi\Omega)\mp\cos(k_{n}L)}\,,
$$
where the $\pm$ sign refers to a number of particles $N$ odd or even 
respectively~\cite{Note1},
$
A_{n,\Omega}=\frac{\sin(k_{n}L/2+\Omega\pi)}{\sin(k_{n}L/2-\Omega\pi)}
$
for $N$ odd and
$
A_{n,\Omega}=-\frac{\cos(k_{n}L/2+\Omega\pi)}{\cos(k_{n}L/2-\Omega\pi)}
$
for $N$ even, while
$\mathcal{N}_n=\sqrt{L \Big( 1+A_{n,\Omega}^2+2A_{n,\Omega}\frac{\sin(k_n L)}{k_n L} \Big) }$
is the normalization factor.

{\it Weak barrier limit} --- In the regime $\lambda\ll 1 $ we determined perturbatively 
the persistent current amplitude, choosing for simplicity $N$ odd. 
The single-particle energy levels $\varepsilon_{n}^{(\pm)}$ are obtained by degenerate perturbation theory 
around the unperturbed parabolas $\epsilon_n=\frac{(2 \pi\hbar)^2}{2M L^2}(\Omega-n)^2$. 
Close to the frustration point $\Omega=1/2$ the pairs of degenerate levels 
are $\epsilon_n$ and $\epsilon_{1-n}$. We immediately get 
\begin{equation}
  \varepsilon_n^{(\pm)} = \frac{(2 \pi\hbar)^2}{2M L^2} \left\{ [\delta \Omega^2+ (n-1/2)^2]
  \pm 2 (n-1/2) \sqrt{\delta \Omega^2+\tilde \lambda_n^2} \right\}
  \label{Eq:tgpert}
\end{equation}
where $\delta \Omega=\Omega-1/2$ and $\tilde \lambda_n=\lambda/(2 n-1)$. 
The TG ground-state energy is given by 
$E_{\rm TG}=\sum_{n=0}^{(N-1)/2} (\varepsilon_n^{(+)}+\varepsilon_n^{(-)})$. 
In this sum the $\pm$ terms in the r.h.s. of Eq.~(\ref{Eq:tgpert}) compensate 
except the highest-energy one~\cite{Note2}, 
yielding Eq.~(4) of the main text with $\lambda_{eff} \simeq \lambda/N$. 
This allows us to choose the short-distance cutoff of the Luttinger theory (see below). 

\section{Details of the soliton-like solution at weak interactions}

We  discuss the soliton-like solution of the GP equation $ \frac{\hbar^2}{2M} ( -i\partial_x - \frac{2\pi}{L}\Omega )^2 \Phi + U_0\delta(x) \Phi + g|\Phi|^2 \Phi = \mu\Phi$. Our approach extends the one of Ref.~\cite{kanamoto1, kanamoto2} to the case where a delta
barrier is present. We first recast the GP equation in dimensionless form by introducing
$\tilde{\Phi} (\theta) = \sqrt{L/2\pi} \Phi(2 \pi x/L)$, $\tilde{\mu} = ML^2 \mu/(2 \pi^2
\hbar^2)$ and $\tilde{g} = gMNL/(\pi \hbar^2)$, and take $\theta \in [0,2 \pi]$. A
parametrization of the condensate wavefunction in density-phase representation $\tilde
\Phi(\theta) = f(\theta) e^{i\phi(\theta)}$ yields
\begin{eqnarray}
  - f'' +f (\phi ')^2 - 2 \Omega f \phi ' + (\Omega^2 - \tilde \mu + \tilde gf^2)f & = & 0\,, \label{equf}\\
  -2 f' \phi ' - f\phi '' + 2\Omega f' & = & 0\,. \label{equphi}
\end{eqnarray}
The effect of the delta barrier is replaced by the cusp condition
$f'(0^+)- f'(0^-) = \lambda f(0)$, where $\lambda = M U_0 L/(\pi \hbar^2)$,
which, assuming a symmetric cusp $f'(0^+) = - f'(0^-)$ and introducing
the density $s=f^2$ reads $s'(0^+) = \lambda s(0)$.
We first integrate Eq.~(\ref{equphi}) to obtain $\phi '$,
\begin{equation}
  \phi' = \frac{C}{f^2} + \Omega\,, \label{angular}
\end{equation}
where $C$ is an integration constant. Substituting this result into Eq.~(\ref{equf})
we get $ -f'' + C^2/f^3 + (\tilde g f^2 - \tilde \mu)f = 0\,$, which, upon integration
and change of variables, yields
$$
s'^2 = -4C^2 + 2 \tilde g s^3 - 4 \tilde \mu s^2 + 4A s\,,
$$
$A$ being an integration constant. Introducing the potential $U(s)=2C^2 -2 As + 2 \tilde
\mu s^2 - \tilde g s^3$, we see that the problem is equivalent to the one of a
classical particle of unitary mass with position $s$ and velocity $s'$ having zero total
energy. Denoting $U(s)= -\tilde g(s-s_1)(s-s_2)(s-s_3)=0$ with $s_1 \le s_2 \le s_3$, an
allowed trajectory is possible in the interval $s_1<s<s_2$ if $A>0$. In the presence of
the barrier, in order to satisfy the the cusp condition, the soliton trajectory starts at
initial position $s_{\rm min}>s_1$. The soliton is then found upon integration,
$$
\int \limits _{s_{\rm min}}^{s(\theta)} \frac{\mbox{d}s'}{\sqrt{-2U(s')}} = \int \limits _{0} ^\theta \mbox{d}\theta ' = \theta
$$
Introducing the change of variable $y^2 = (s-s_1)/(s_2-s_1)$, the integral corresponds
to the Jacobi elliptic function $\mbox{sn}(u|m)$ with $m = (s_2-s_1)/(s_3-s_1)$.
By imposing periodic boundary conditions for the condensate density ({\it i.e.}, requiring
that at half-period the soliton solution reaches its maximum value $s=s_2$),
we find $\pi \sqrt{\tilde g(s_3-s_1)/2} + \alpha = K$ and
\begin{equation}
  s(\theta) = s_1 + (s_2-s_1) \mbox{sn}^2[(K-\alpha)\theta /\pi + \alpha]. \label{formbarfin}
\end{equation}
Here $\alpha = F[\arcsin(\sqrt{(s_{min} - s_1)/(s_2-s_1)})|m] \label{alpha}$
with $F[\phi|m]$ being the incomplete elliptic integral of the first kind,
and $K$ is the corresponding complete elliptic integral.

Imposing the normalization condition $\int_{0}^{2\pi} d\theta s(\theta)=1 $ we obtain
$$
2 \pi s_1 + 4 (K-\alpha)(K-\alpha-E+\alpha')/(\pi \tilde g) =1,
$$
where $E[\phi|m]$ and $E$ are respectively the incomplete and complete elliptic integrals
of the second kind, and $ \alpha' = E[\arcsin(\sqrt{(s_{min} - s_1)/(s_2-s_1)})|m] \label{alphapr}\,.$
Substituting Eq.~(\ref{formbarfin}) for $s$ in equation $s'^2 = -2U(s)$ and equating
the terms with the same power of sn, yields the following parameter identification:
\begin{eqnarray}
  \tilde \mu & = & 3 \tilde gs_1/2 + (1+m) [(K-\alpha)/\pi ]^2 \, ; \label{mubar}\\
  A & = & 2\tilde \mu s_1 - 3\tilde gs_1^2/2 - 2 m [(K-\alpha)/\pi ]^4/\tilde g \, ; \label{Abar} \\
  C^2 & = & A s_1 - \tilde \mu s_1^2 + \tilde gs_1^3/2 \, . \label{C1bar}
\end{eqnarray}
Finally, the cusp condition yields an equation for $s_{\rm min}$,
\begin{eqnarray}
  {\lambda s_{\rm min} = \sqrt{-2U(s_{\rm min})}.}\label{smin}
\end{eqnarray}
Equations~(\ref{mubar} -- \ref{smin}) form a coupled set of equations
which can be solved as a function of $m$ and {$s_{\rm min}$}.

Using Eqs.~(\ref{angular}) and~(\ref{formbarfin}), we then obtain the solution
for the phase of the soliton solution:
\begin{equation}
  \phi(\theta)=C(\Pi[(s_1-s_2)/s_1;{(K-\alpha)\theta/\pi+K}|m] - \Pi[(s_1-s_2)/s_1;\alpha|m])/(K-\alpha) s_1+\Omega(\theta+\pi)\,,
  \label{fase}
\end{equation}
where $\Pi[n;u|m]$ is the incomplete elliptic integral of the third kind.\\
Imposing $2\pi$-periodicity of the condensate wave function, integration
of Eq.~(\ref{angular}) yields $2\pi n = \int \limits _{-\pi}^\pi d\theta \frac{C}{s} + 2\pi \Omega$,
hence
\begin{equation}
  2 \pi (n-\Omega)/C = 2\pi (\Pi[(s_1-s_2)/s_1;K|m] - \Pi[(s_1-s_2)/s_1;\alpha|m])/((K-\alpha) s_1)\,.
  \label{C2bar}
\end{equation}

\begin{figure}[ht]
  \centering
  \includegraphics[width=0.3\textwidth]{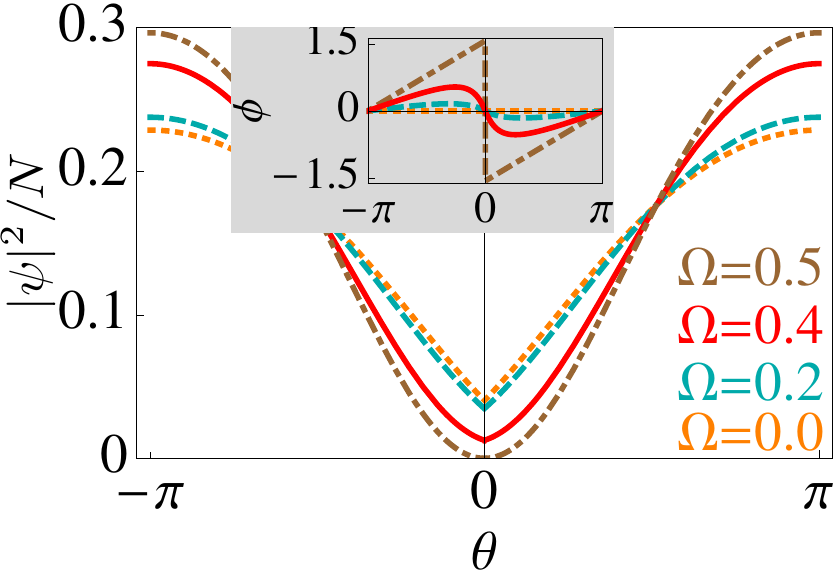} \includegraphics[width=0.3\textwidth]{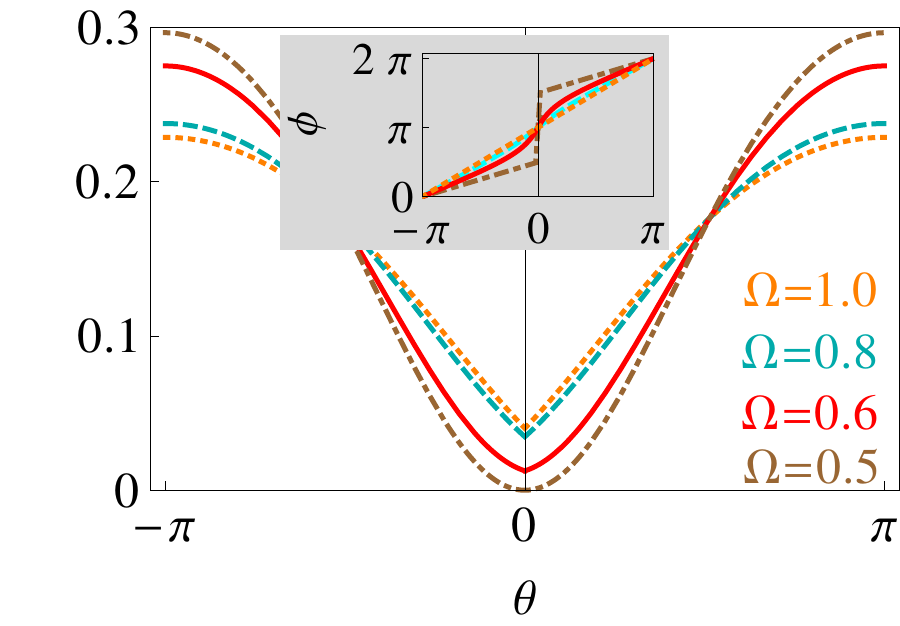} \includegraphics[width=0.3\textwidth]{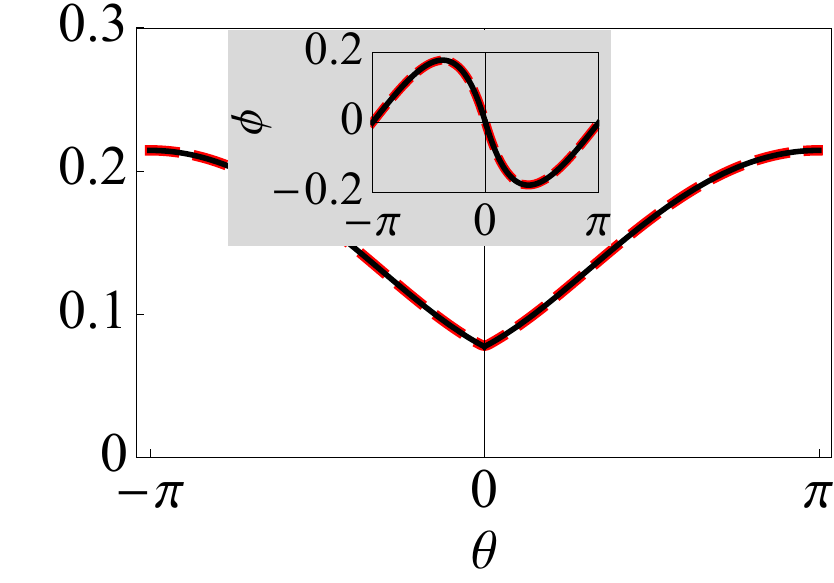}
  \caption{Left and central panel, soliton solution for the density (main figure) and phase (inset) as a function of the spatial coordinate along the ring for various values of the reduced Coriolis flux $\Omega$ as indicated in each figure, at fixed dimensionless barrier height $\lambda=1.9$, $N=18$ and $\tilde g=1$. 
Right panel, comparison of the analytical soliton solution of the GP equation (black solid line),
with the numerical solution obtained via imaginary-time integration (red dashed line) taking $N=18$, $\Omega=0.4$, $\lambda=0.5$,
and $\tilde g=1$. The numerical ground state energy
in units of $2 \pi^2 \hbar^2/M L^2$ is $\tilde E^{\rm num}_{\rm GS}={0.2965(3)}$
to be compared with the one of the soliton solution {$\tilde E^{\rm sol}_{\rm GS}={0.2966}$}.}
  \label{agreement}
\end{figure}

Fig.~\ref{agreement} shows the soliton density and phase at fixed barrier strength at varying Coriolis flux $\Omega$, both for the $n=0$ case of no winding, and of one quantum of angular momentum, ie $n=1$. 

In summary, we find the soliton solution according to the following strategy:
for given values of the interaction constant $\tilde g$, the barrier strength $\lambda$
and the Coriolis flux $\Omega$, we express $s_1$, $s_{2}$ and $s_3$
(hence $\tilde \mu$, $A$, and $C$) as a function of $m$ and $s_{\rm min}$,
then solve simultaneously Eq.~(\ref{smin}) and the one obtained after equating
Eqs.~(\ref{C2bar}) and~(\ref{C1bar}). This uniquely fixes $m$ and $s_{\rm min}$,
and hence the entire soliton solution.

By performing imaginary-time numerical integration of the GP equation,
we have checked that the analytical soliton solution given by
Eqs.~(\ref{formbarfin}),~(\ref{fase}) coincides with the numerical ground state,
see again Fig.~\ref{agreement}.

\newpage
\section{Details of the Luttinger-liquid solution}

We outline the derivation yielding Eqs.(4) and (5) of the main text.

In the weak barrier case, we start from the mode expansion for the density
and phase fields for the uniform ring,
\begin{eqnarray}
  \theta(x) & = & \theta_{0} + \frac{1}{2}\sum_{q\neq 0}\left|\frac{2\pi K}{qL} \right|^{1/2}[e^{iqx}b_{q}+e^{-iqx}b_{q}^{\dagger}] \, , \\
  \phi(x) & = & \phi_{0} + \frac{2\pi x}{L}(J-\Omega)+\frac{1}{2}\sum_{q\neq 0}\left|\frac{2\pi}{qLK} \right|^{1/2} \operatorname{sgn}(q)[e^{iqx}b_{q}+e^{-iqx}b_{q}^{\dagger}] \, , 
\end{eqnarray}
where $q=2\pi j/L$ with $j$ integer, $J$ is the angular momentum operator, $K$ is here the Luttinger parameter
and the following commutation relations hold:
$[b_{q},b_{q'}^{\dagger}]=\delta_{q,q'}$; $[J,e^{-2i\theta_{0}}]=e^{-2i\theta_{0}}$.
The latter property implies that the zero-mode $\theta_0$ acts as a (normalized) raising operator
for the states $|J\rangle$ of given angular momentum: $e^{-2i\theta_{0}}|J\rangle=|J+1\rangle$.

The lowest-order relevant term in the barrier Hamiltonian induces transitions of
one quantum of angular momentum~\cite{Note3}
due to the zero-mode part in $\theta(x)\equiv\theta_{0}+\delta\theta(x)$, {\it i.e.}
$$
\mathcal{H}_{b}\sim 2 U_{0}n_{0}\cos(2\theta(0))=n_0 U_0 \sum_J |J-1\rangle\langle J| e^{2 i \delta \theta(0)}
+ |J\rangle\langle J+1| e^{-2 i \delta \theta(0)}.
$$

The calculation is performed in two steps. First, using the mode expansion of the
fields and averaging the total Hamiltonian over the nonzero modes, we obtain an
effective Hamiltonian for the angular momentum operator~\cite{GogPro94}
\begin{equation}
  \mathcal{H}_{J}=E_{0}(J-\Omega)^2+n_{0}U_{\rm eff}\sum_{J}|J+1\rangle\langle J|+ {\rm H.c.}\,.
\end{equation}
The effective barrier strength $U_{\rm eff}$ is obtained by averaging over
the density fluctuations $U_{\rm eff}=U_0 \langle e^{\pm i 2 \delta \theta (0) }\rangle = U_{0}(d/L)^K$,
$d$ being a short-distance cutoff of the order of the interparticle distance.
By choosing $d=K/n_0$, this expression coincides with the exact TG result
at $K=1$ (see above) and takes into account the shrinking
of the linear region of the excitation spectrum once interactions are decreased
away from the TG point.

In a second step, we perform degenerate perturbation theory between states of given
angular momentum, obtaining
$E(\Omega)/N=E_0 [\delta\Omega^{2}-\sqrt{\delta\Omega^{2}+\lambda_{\rm eff}^{2}/4\pi^{2}}]$,
with $E_0=2 \pi^2 \hbar^2/ML^2$, $\delta\Omega=\Omega-J-1/2$,
and $\lambda_{\rm eff} = MU_{\rm eff}L/\pi\hbar^{2}$.
Upon using the thermodynamic relation, Eq.~(2) of the main text
one readily obtains Eq.~(4) of the main text.

In the strong barrier or weak-tunnel limit the appropriate mode expansion
is the one with open boundary conditions,
\begin{eqnarray}
  \theta(x) & = & \theta_{0} + i\sum_{q>0}\left(\frac{\pi K}{qL} \right)^{1/2}\sin(qx)[b_{q}-b_{q}^{\dagger}]\,,\\
  \phi(x) & = & \phi_{0} + \sum_{q> 0}\left(\frac{\pi}{qLK} \right)^{1/2}\cos(qx)[b_{q}+b_{q}^{\dagger}] \, ,
\end{eqnarray}
where $q=j\pi/L$ with $j$ integer. Note that no circulation is allowed
in the limit of infinitely large barrier, and hence the angular momentum $J$
does not enter here. As in the weak-barrier case, we average the barrier-contribution
of the Hamiltonian over the fluctuation modes obtaining the $\Omega$-dependent part of the energy,
$E(\Omega)/N=- 2 (t/L) \langle \cos(\phi(L)-\phi(0)+ 2 \pi \Omega)\rangle
= -2(t/L)(d/L)^{1/K}\cos(2\pi\Omega)$, where we have made the same choice
for the short-distance cut-off length $d$ as in the weak-barrier case above.
This readily leads to Eq.(5) of the main text.

\begin{figure}[ht]
  \centering
  \includegraphics[width=0.3\textwidth]{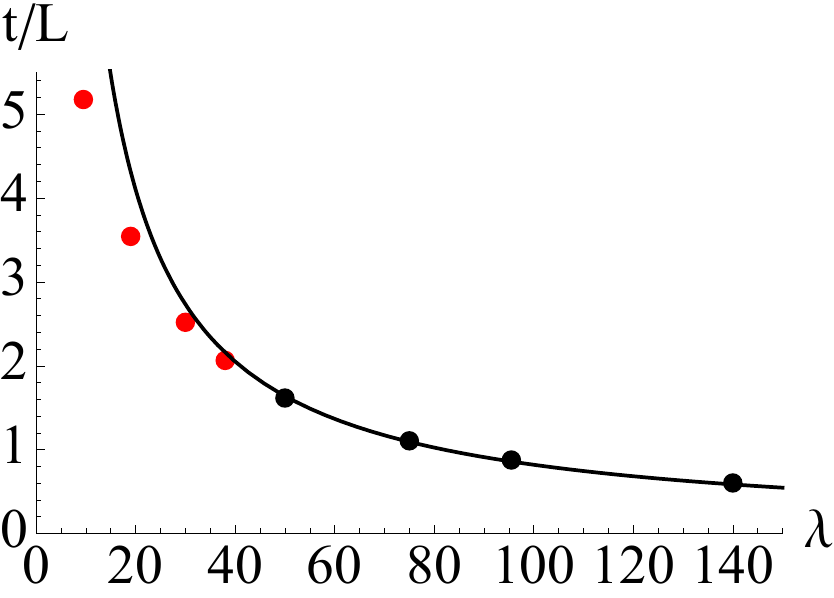}
  \caption{Tunneling amplitude $t$ {\it vs.} barrier strength $\lambda$ in the TG limit for $N=18$.
    For the red dots, the value of $\lambda$ is too weak for the tunneling approximation
    to hold, and a deviation from the hyperbolic behavior is observed.}
  \label{iperbole}
\end{figure}

The relation between tunneling amplitude $t$ and dimensionless barrier strength
$\lambda=MU_{0}L/\pi\hbar^2$ at given interparticle interactions is obtained using the result~\cite{weiss},
\begin{equation}
  t/L = \Gamma(1+K) \, \Gamma(1+1/K)^{K} \left( \hbar\omega_{c}\right)^{1+K} (U_{0}/L)^{-K},
  \label{eq:duality}
\end{equation}
where the model-dependent cut-off frequency $\omega_c$ is determined using
the exact TG result in the case $K=1$. In essence, for large values of the barrier strength
we perform exact calculations of the current $I$ vs Coriolis flux $\Omega$,
then use Eq.~(5) of the main text as fitting function to extract the value
of the tunnel amplitude $t$. The resulting dependence of $t$ on the barrier strength
is shown in Fig.~\ref{iperbole}. For sufficiently large values of the barrier strength
({\it i.e.}, $\lambda\gtrsim 50$) a very good agreement is found with the hyperbolic
law~(\ref{eq:duality}), allowing therefore to extract $\omega_c$.

\vspace{-0.5cm}
\section{Details of the matrix-product-state numerical approach at intermediate interactions}

In order to tackle the ground-state many-body problem for intermediate interactions, 
we resorted to numerical simulations based on a discretized version of 
the Hamiltonian~(1) in the main text.
This can be written in terms of a Bose-Hubbard model for a 1D chain of $N_s$ lattice sites
with periodic boundary conditions (PBC) [see Hamiltonian~(6) in the main text].

By taking the continuum limit of Eq.~(6) in the main text, we link the parameters $t_{\rm BH}, U_{\rm BH}$
and $\beta$ of the Bose-Hubbard model to the parameters $M, g$ and $U_{0}$
of the continuum Hamiltonian~(1) in the main text.
Specifically we have $\hbar^{2}/2M = t_{\rm BH}a^2$, $g = U_{\rm BH} a$ and $U_0 = a\beta$,
where $a$ is the lattice spacing of the discrete model.

Our simulations are based on the Density Matrix Renormalization Group (DMRG) 
method, (see, {\it e.g.}, Ref.~\cite{Schollwock}).
Specifically, we have used a Matrix Product State (MPS) representation
of the many-body wavefunction and performed a variational minimization
of the energy cost function, ruled by Hamiltonian~(6) in the main text, site by site.
We controlled the accuracy of the simulation by adjusting the size $m$
of each matrix (bond link) composing the MPS.
It is worth mentioning that special care has to be taken
for PBC systems with a DMRG-based algorithm,
since it performs a factor $m^2$ worse than for open-ended systems~\cite{Verstraete,Murg}.
Here we adopted an improved method based on the controlled
factorization procedure for long products of MPS transfer matrices,
which takes into account only $p \ll m^2$ singular values, without
compromising the accuracy and reducing the computational
effort~\cite{Pippan, PBC_stiff, Weyrauch}.
This is justified by the fact that, for large chains,
the local physics of the system is weakly affected by the properties
of the boundaries.
In a similar fashion, we approximated the effective Hamiltonian
on the MPS basis by expanding it via a singular value decomposition,
keeping only the contributions associated to its $s \sim p$
largest eigenvalues~\cite{PBC_stiff}.
Unfortunately the implementation of any symmetry
in the MPS wavefunction, such as, in our case, the one accounting 
for the conservation of the total number of bosons, is 
less trivial with PBC rather than with OBC~\cite{Note4}.
This forces us to work in the grand canonical ensemble and fix
an average number of particles by suitably tuning the chemical potential.

In our simulations we have chosen {\bf $N_s = 120$} sites
and fixed the chemical potential $\mu = \mu(t_{\rm BH},U_{\rm BH})$ in such a way
to have $N \sim 18$ particles in the system.
A fine-tuning of $\mu$ in order to have a well defined number of particles
has been performed in absence of any barrier, in the case when the
system is translationally invariant (the effect of a finite $\beta$
is that of smearing such accuracy in $N$---see below).
We thus kept an average low filling
$\langle n_j \rangle \lesssim 0.15$ (here $n_j = b^\dagger_j b_j$ counts
the number of bosons on site $j$), mimicking the continuum limit
of Eq.(2) of the main text and minimizing lattice effects.
The simulations presented in Fig.~(2) and Fig.~(3) of the main text
have been performed with $m = 20$, while we fixed $p = s$ up to $120$.
We also cut the maximum allowed occupation per site to
$n_{\max}$ bosons, going from $n_{\max} = 1$ in the TG regime, to
$n_{\max} = 4$ for $U_{\rm BH} = 0.1$ (the lowest chosen value for $U_{\rm BH}$,
in units of $t_{\rm BH}$).
\begin{figure}[t]
  \centering
  \includegraphics[width=0.45\textwidth]{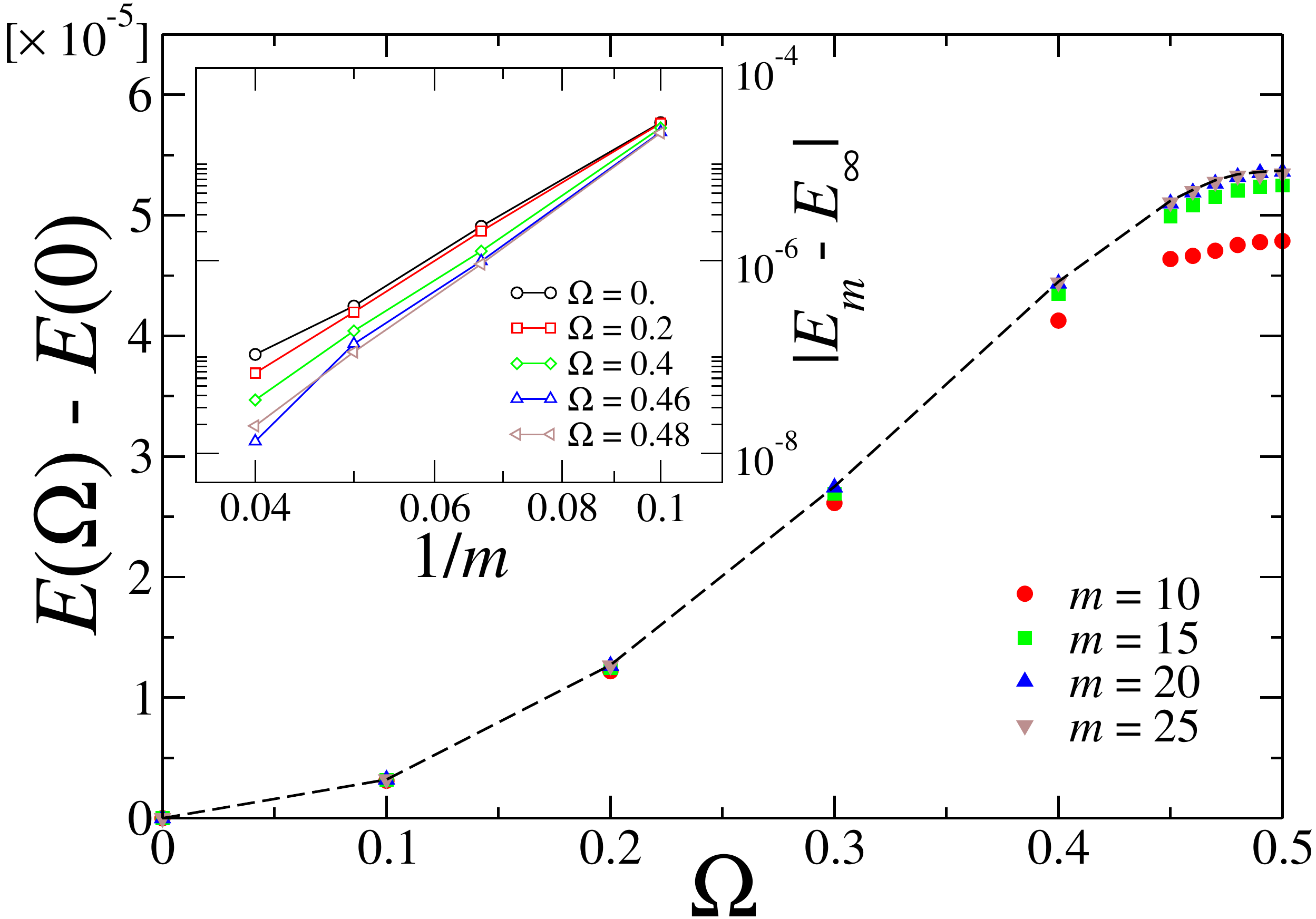}
  \caption{Ground-state energy {per lattice site, in units of $t_{\rm BH}$}, for the system described by the Hamiltonian~(6) in the main text,
    as a function of the phase twist $\Omega$ and for different values of the bond-link
    dimension $m$ (here we fixed $p = s = 4 m$).
    The dashed line is obtained by extrapolating the values of the energies for $m \to \infty$,
    according to a power-law fit of the numerical data (symbols): $E_m \sim E_{\infty} + A \, m^{-B}$.
    Here we simulated a Bose-Hubbard chain of length $N_{s} = 120$, with $U_{\rm BH} = 10$,
    $\mu = -1.81$, $\beta = 1$ (parameters are expressed in units of $t_{\rm BH}$).
    We also chose $n_{\rm max} = 2$, checking that this is sufficient with this value of interaction.
    Inset: Scaling of the energies with $m$, for fixed values of $\Omega$.}
  \label{Energy_m_scal}
\end{figure}

We carefully checked that $m \simeq 20$ is already sufficient to get reliable
results for the particle current $I(\Omega) \propto -\partial_\Omega E$,
where $E(\Omega)$ is the ground-state energy.
An example of the typical performances of our MPS code as varying 
the bond-link dimension $m$ is provided in Fig.~\ref{Energy_m_scal}, 
where we explicitly show the dependence of the ground-state energy $E$ 
as a function of the phase twist $\Omega$.
For fixed $\Omega$, we found a dependence of such energy on $m$
consistent with a power-law behavior: $E_m \sim E_{\infty} + A \, m^{-B}$, 
from which we could extrapolate the asymptotic value $E_{\infty}$ (inset).
We point out that the bond link values considered in this work
are considerably smaller than those used in typical DMRG simulations 
for open-ended chains,
where $m_{\rm obc} \sim 500$ could be easily reached.
Nonetheless they are sufficient to reach an accuracy of the order $10^{-7}$
in terms of absolute values of energies, thus leading to an error in the 
extrapolation of the particle current $I(\Omega)$ of the order of $1 \%$,
that is barely visible on the scale of Fig.~\ref{Energy_m_scal}.
A further and most important source of inaccuracies comes from the determination 
of the average number of particles $N$, which in some cases (especially for 
small interactions) becomes less accurate when varying the barrier strength.
As a combination of these two sources of errors, the bars in Fig.~3 of the main text
have been computed from point to point.

We found a very similar scenario for all the interaction regimes
considered in this work.


\begin{thebibliography}{10}
\bibitem{mps}
  The MPS approach generally performs worse in the weakly interacting regime,
  where single-particle wavefunctions are delocalized across the lattice.
  Moreover the soft-core nature of bosons increases the inaccuracies
  due to the cutoff in the maximum on-site occupancy~\cite{SM}.
\end{thebibliography}
\end{document}